\RequirePackage{color}  
\documentclass{PoS}
\usepackage{amsmath}
\usepackage{booktabs}

\usepackage[dvipsnames]{xcolor}
\definecolor{dkgreen}{rgb}{0.0,0.4,0.0}
\definecolor{red}{rgb}{0.8,0.0,0.0}
\definecolor{darkblue}{rgb}{0.0,0.1,0.7}
\newcommand{\green}[1]{{\color{dkgreen} #1}}
\newcommand{\red}[1]{{\color{red} #1}}
\newcommand{\blue}[1]{{\color{darkblue} #1}}
\usepackage{bm}
\graphicspath{{figs/}}
 
\title{Fast Partitioning of Pauli Strings into Commuting Families for Expectation Value Measurements of Dense Operators}

\ShortTitle{Fast Pauli Partitioning}

\author{
\speaker{Nouman Butt},
\speaker{Andrew Lytle},
Ben Reggio, and
Patrick Draper
\newline
\\
Department of Physics and Illinois Center for Advanced Studies of the Universe, \\University of Illinois, Urbana, Illinois, 61801, USA\\
}

\abstract{
The cost of measuring quantum expectation values of an operator can be reduced by grouping the Pauli string ($SU(2)$ tensor product) decomposition  of the operator into maximally commuting sets.  We detail an algorithm, presented in [1], to partition the full set of $m$-qubit Pauli strings into the minimal number of commuting families, and benchmark the performance with dense Hamiltonians on IBM hardware. Here we also compare how our method scales compared to graph-theoretic techniques for the generally commuting case.
}

\FullConference{%
The 40th International Symposium on Lattice Field Theory,\\
July 31st -- August 4th, 2023,\\
Fermilab, Batavia, Illinois, USA}

\begin{document}
\section{Introduction}
The Pauli strings $P_i$ appearing in the decomposition of an $m$-qubit operator $H$,
\begin{equation}
H = \sum^{4^{m}}_{i=1} c_{i}P_{i}
\end{equation}
where $P_i$ is a tensor product of Pauli matrices, for example
\begin{equation}
   P = \sigma_x\otimes \mathbf{1} \otimes \mathbf{1} \otimes \sigma_y \otimes\dotsc \otimes \sigma_z \equiv XIIY \dotsc Z \,,
\end{equation}
can be grouped into commuting families,  reducing the number of quantum circuits needed to measure the expectation value of the operator. We detail an algorithm to completely partition the full set of Pauli strings acting on any number of qubits into the minimal number of sets of commuting families, and we provide python code to perform  the partitioning. The partitioning method scales linearly with the size of the set of Pauli strings and it naturally provides a fast method of diagonalizing the commuting families with quantum gates. We provide a package that integrates the partitioning into Qiskit, and use this to benchmark the algorithm with dense Hamiltonians, such as those that arise in matrix quantum mechanics models, on IBM hardware. We demonstrate computational speedups  close to the theoretical limit of $(3/2)^m$ relative to qubit-wise commuting groupings, for $m=2,\dotsc,6$ qubits~\cite{Reggio:2023fue}.
   
The cost of a quantum computation depends on several aspects of the computation, including the number of required quantum circuits, the depth of the circuits, and the number of times the same circuits have to be run in order to achieve a level of confidence in the results. For computations involving expectation value measurements, e.g.\ variational quantum eigensolver (VQE) problems, the na\"ive approach for a generic operator  produces  $\mathcal{O}(4^m)$ circuits for $m$ qubits (one for each Pauli string in the operator decomposition).   In the NISQ era, the capacity to share the computational burden between classical and quantum computers in an optimal way will be crucial. This is a classical problem, the solution of which can be used to reduce the number of circuits needed to measure an expectation value on a quantum device. An optimal solution partitions all Pauli strings into $2^m +1$ sets (families) where each set has size $2^m -1 $. This partition reduces the number of circuits from $4^m$ ($3^m$), in the  na\"ive (qubit-wise commuting) case, down to $2^m+1$. 
 
\section{Properties of Pauli Strings}
A family is defined as a maximally commuting set of Pauli strings. All families have the same size $2^m -1 $ and can be generated from $m$ generating strings~\cite{Reggio:2023fue}. These generating strings are mutually independent: none of them can be written as a product of other strings. In other words we only need $m$ generating strings to characterize a family since all other strings in the family are all possible products of these $m$ generating strings. The two canonical families, namely $z$ family with strings of characters $I$s and $Z$s and $x$ family with characters $I$s and $X$s serves as generators of an optimal partition. The full set of Pauli strings can be constructed by taking products of strings in the canonical families as shown in the  $2$-qubit table below. The color scheme illustrates different families in the $2$-qubit case. In the next section we explain the algorithm to obtain these families.
  \begin{table}[ht]
    \centering
    \begin{tabular}{c|ccc}
         & $X \otimes 1 $ &$1 \otimes X $ & $X \otimes X $\\
         \hline
        $Z \otimes 1 $ &  \green{$Y \otimes 1$} & \red{$Z \otimes X$} & \blue{$Y \otimes X$} \\
        $1 \otimes Z$ & \blue{$X \otimes Z$} & \green{$1 \otimes Y$} & \red{$X \otimes Y$} \\
        $Z \otimes Z $ & \red{$Y \otimes Z$} & \blue{$Z \otimes Y$} & \green{$Y \otimes Y$}
    \end{tabular}
\end{table}   

\section{Properties of matrix A}
 The problem of finding optimal partition for full set of Paulis has been investigated many times in the past. Notably Jena developed an  approach which reduces the partitioning to finding a set of $Z_2$-valued, $m\times m$ matrices $A_{i}$~\cite{Jena2019}. 
 The canonical $x$ and $z$ families can be converted into a $Z_2$-valued vector space $\mathcal{V}$ with the generating strings playing the role of basis vectors for $\mathcal{V}$. The $i$-th family that can be generated from the canonical $x$ and $z$ families has a corresponding generator matrix $A_{i}$ which encodes the commutativity of the strings in that family and furnishes a unique permutation on the $x$ family by transforming  its basis vectors into a new set of basis vectors.
 
 These matrices $A_{i}$ are  symmetric, and can be represented as powers of a single matrix $A$. This set of matrices  forms a Singer cycle $A_i \in \{A, A^2 , A^3, ....,A^{N}=A\}$. The matrix $A$ has a period given by $A^{2^m-1}=1$. On the generating strings (basis vectors $v_{i } \in \mathcal{V}$) the matrix $A$ acts as a permutation:
\begin{equation}
 Av_{i} = v_{P(i)}  
\end{equation} 
Every successive matrix $A_i (= A^{i})$ realizes a distinct permutation on the $x$ family generating strings and provides a distinct new set of generating strings for the $x$ family. Using this new set we obtain generating strings for a new family by taking products with $z$ family's generating strings. In order to avoid redundancies in the solution we need to make sure that every matrix $A_{i}$ should have the right characteristics: 
\begin{itemize}
\item $A_{i}$ is symmetric
\item $A_i - A_j$ for $ \forall j\neq i$ is invertible 
\end{itemize}  
In order to generate the set of matrices $A_{i}$ we use the matrix representation of the Galois field $GF(2^m)$. This representation leads to a set of matrices $\{C, C^2, ...C^{N-1}\}$ which has the Singer cycle property~\cite{Jena2019}. 
However these matrices requires symmetrization which is done using a separate method~\cite{1998.0216}.

\section{Diagonalizing to $z$ family}
 With a solution of $2^m +1 $ families, we need to run only $ O(2^m)$ circuits rather than $4^m$. The computational basis which is used for measurement is the eigen-basis for the $z$ family. For each family we need to find a unitary transformation that can transform the strings in the family to strings in $z$ family modulo an overall sign.  Generating strings $\{x_m\}$ of the canonical $x$ family can be used to obtain the unitary transformation via $U=exp(i\frac{\pi}{4}\sum_m x_m )$. The overall sign can be evaluated by keeping track of the sign change accumulated for each generating string. Surprisingly the set of matrices $\{A,A^2,A^3,...\}$ can be used to find the diagonalizing strings $\{x_m\}:$.
  
 For the $i-$th family given by $A^{i}$ the diagonalizing strings can be found via computing $(A^{i})^{\frac{N}{2}}$. 
 \begin{equation}
    (A^i)^{N/2} = \left\{ \begin{array}{cc}
        A^{i/2} & \text{if } i \text{ mod } 2 = 0 \\
        A^{\frac{N+i-1}{2}} & \text{if } i \text{ mod } 2 = 1
    \end{array} \right\}.
    \label{AiN2}
\end{equation}
The set of generating (diagonalizing) strings $\{x_m\}$ are obtained from the action of $(A^i)^{N/2}$~\cite{Reggio:2023fue}. This unitary transforms the given family to the canonical $z$ family. However for simultaneous measurement of all the strings in the family, this unitary transformation leads to an additional circuit depth which is approximately quadratic in the number of qubits. In QWC (qubit-wise commuting) families the measurement circuit depth only increases by a unit. In the section below we show that this additional circuit depth marginally affects the overall runtime.

\section{Integrating into QISKIT}
We developed a python package for generating the optimal solution and the diagonalizing circuits~\cite{psfam}. We also developed a {\tt QISKIT} extension {\tt dense\_ev}~\cite{dense-ev} which contain two classes. The first class {\tt DenseGrouper} works as an analog of native {\tt QISKIT} class {\tt AbelianGrouper} (which generates qubit-wise commuting solutions) and the second class {\tt DensePauliExpectation} builds upon the native {\tt QISKIT} {\tt PauliExpectation} and contains the method to compute expectation values on hardware and quantum simulators. Both packages are publicly available.

\section{Computational Cost}
We compare the computational cost of qubit-wise commuting (QWC) vs.\ optimal grouping using a simple model for the runtime of a single circuit, $\tau = \tau_{over} + \tau_{circ}(D) $, where $\tau_{over}$ is the circuit overhead time and $\tau_{circ} $ depends on the depth $D$ of the circuit needed to generate the state $| \psi \rangle$ in the desired expectation value $\langle \psi | H | \psi \rangle$. We assume that the runtime $t_{circ}(D)$ is roughly linear in the circuit depth.
For QWC groups, the post-state rotation circuit to transform to the computational basis has depth 1, while for groups based on optimal grouping the rotation circuit has depth proportional to $m^2$.
For a prepared $m$-qubit state of depth $D$ we have
\begin{equation} 
\frac{\tau_{\text{QWC}}}{\tau_{\text{optimal}}} = 
\frac{3^m \Bigl[\tau_{\text{over}}+\tau_{\text{circ}}(D+1)\Bigr]}{(2^m+1) \Bigl[ \tau_{\text{over}}+\tau_{\text{circ}}(D+am^2) \Bigr]} \,.
\end{equation} 
 If the circuit overhead is much greater than the circuit runtimes ($ \tau_{over} >> \tau_{\text{circ}}$), or the state circuit depth is much greater than the average diagonalization circuit depth ($D >> am^2$), the runtime improvement will be close to the ideal $(3/2)^m$.

\section{Numerical Results}       
\begin{figure}[h!]
    \centering
    \includegraphics[width=0.45\linewidth]{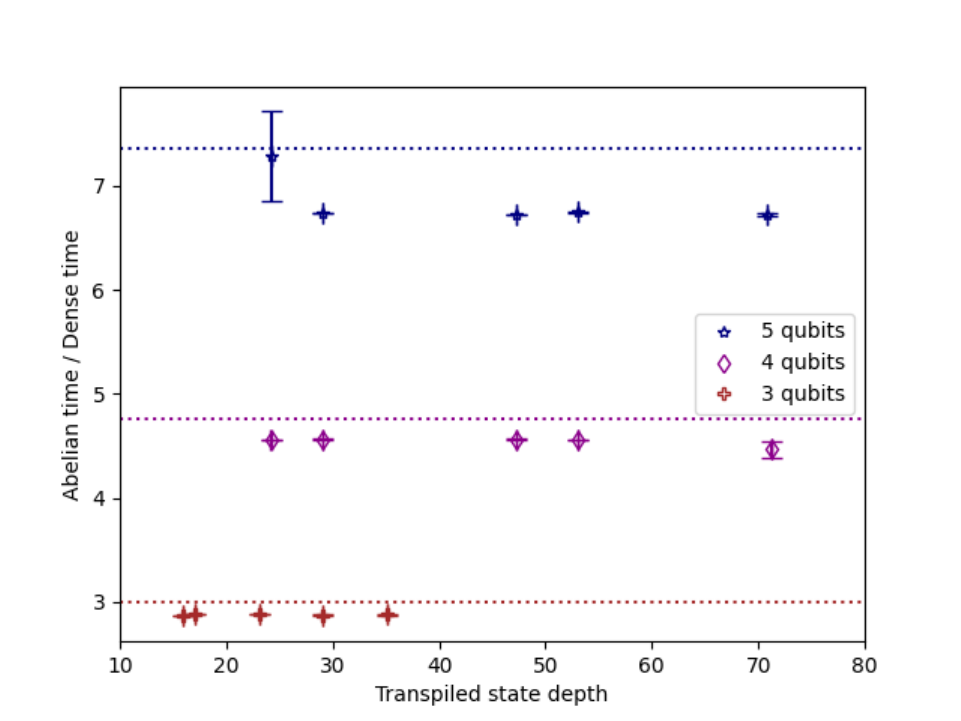}
    \caption{The ratio of computation times to run circuits needed for expectation value measurement using QWC (Abelian) and optimal (dense) grouping, on {\tt ibmq\_quito}.}
    \label{fig:quito}   
\end{figure}
 In  Fig.~\ref{fig:quito} we  show the ratios of the computational times between different methods. The circuits ran on {\tt ibmq\_quito} using both dense (optimal) and abelian (QWC) grouping methods for 3 to 5 qubits. The ideal speedup factor is the ratio of the number of circuits, $\frac{3^m}{2^m + 1}$ shown by dotted lines. The states measured are constructed using {\tt EfficientSU2}, and the {\tt reps} parameter is varied from 1 to 5 to show the effects of increased circuit depth.

\begin{figure}[h!]
    \centering
    \includegraphics[width = 0.45\linewidth]{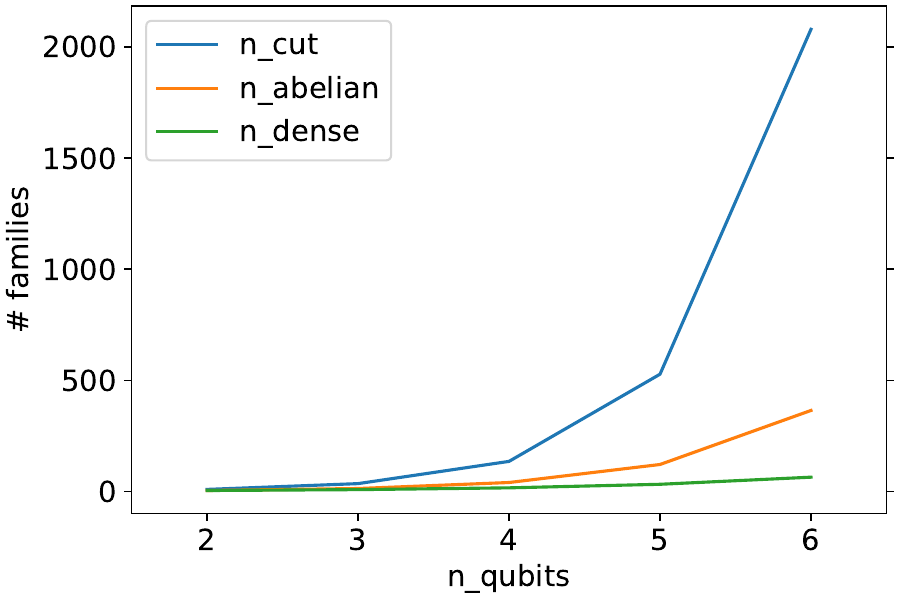}~~
    \includegraphics[width = 0.45\linewidth]{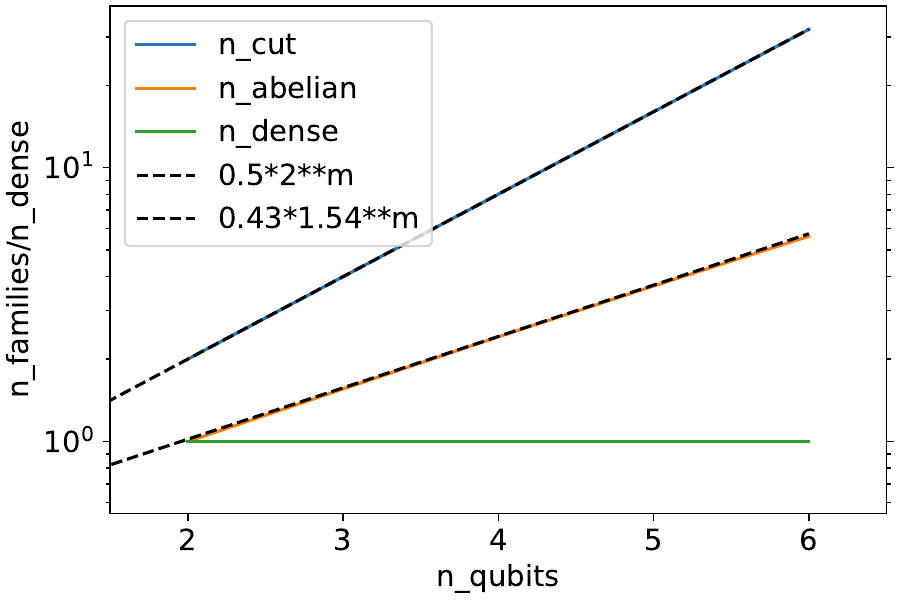}
    \caption{(Left) Number of families generated by different grouping methods for the Femtouniverse Hamiltonian. (Right) Same data as the left figure but expressed as a ratio to the number of families generated by the dense method.}
    \label{fig:nfamilies}
\end{figure}

On the left of Fig.~\ref{fig:nfamilies} we show the number of family groupings generated by different grouping methods for the $A_1^+(g=0.8)$ femtouniverse Hamiltonian~\cite{Butt:2022xyn}, as a function of number of qubits $m$. We compare the na\"ive decomposition into individual Pauli strings, the {\tt AbelianGrouper}, and the dense grouping. On the right we have the same data but plotted as a ratio to the number of families from the dense grouping ($2^m+1$),
showing the improvement factor of the dense grouping compared to measuring individual Pauli strings (blue) and grouping generated by {\tt AbelianGrouper} (orange). The dotted lines give an indication of the exponential improvement observed using the dense vs.\ other methods. 

\section{Comparison with graph theoretic methods}
\begin{figure}
    \centering
    \includegraphics[width=0.49\textwidth]{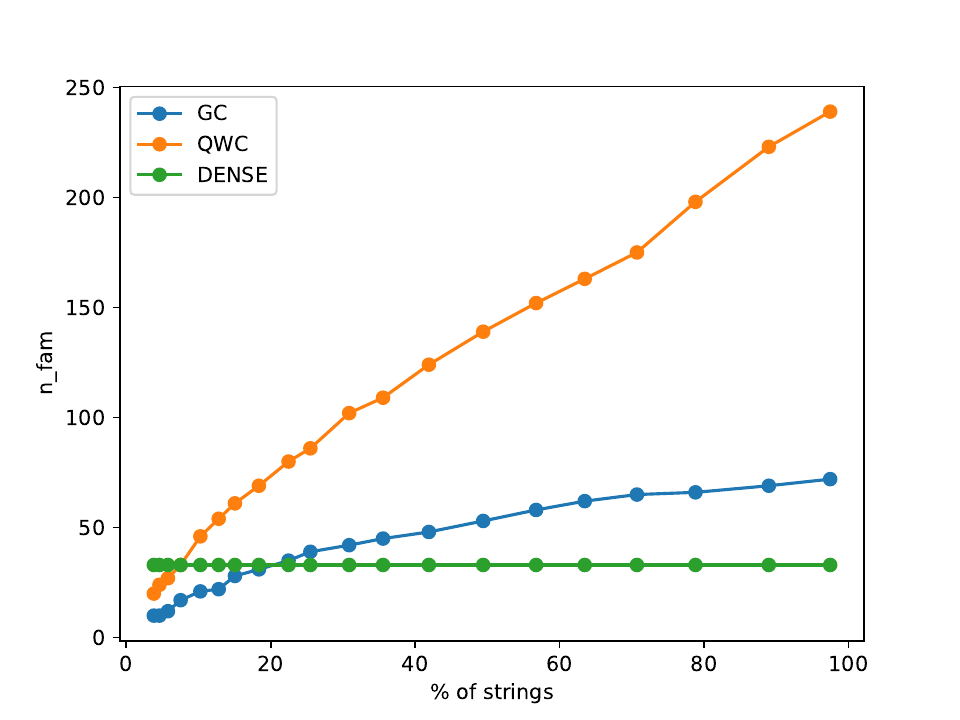}
    \includegraphics[width=0.49\textwidth]{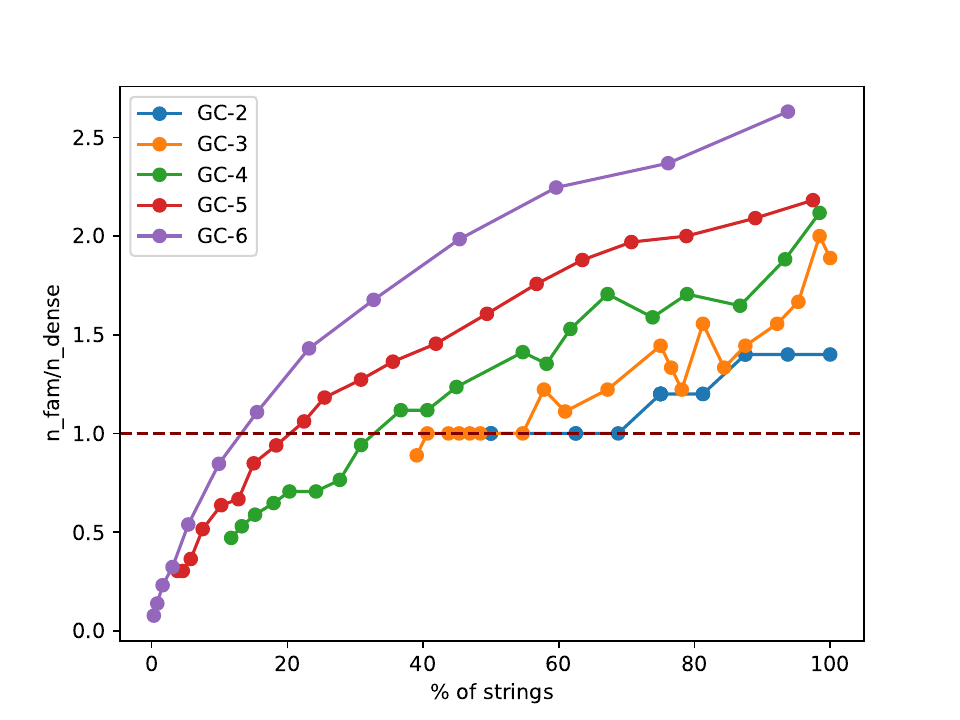}
    \caption{(Left) Number of family groupings generated by different methods, starting from a random Hermitian operator, with $m=5$. The number of Pauli strings in the original operator is reduced by applying a numerical cut on the string coefficients, and the percentage of the original $4^m$ strings remaining after the cut is plotted on the $x$-axis.  (Right) Ratio of number of families produced by the GC method to that produced by the DENSE method, for $m \in [2,6]$, as a function of the percentage of the original $4^m$ strings present in the operator after a numerical cut on the coefficients.}
    \label{fig:nfam}
\end{figure}
The problem of partitioning a set of Pauli strings into a minimal number of commuting sets may be re-expressed as a graph theory problem~\cite{gokhale2019minimizing,1907.07859,1907.03358,1907.09386,1907.09040}, where the Pauli strings in an operator represent nodes of the graph and presence of an edge between nodes represents whether the strings commute (or anti-commute, depending on the problem formulation)\footnote{Every clique on the commutation graph is dual to an independent set on the complement (anti-commutation) graph.}. 
In~\cite{Reggio:2023fue}, we compared our algorithm with graph methods based on qubit-wise commuting (QWC) groups, and here we extend this comparison to methods based on generally commuting (GC) groups.

For fully dense operators (all $4^m$ Pauli strings present), the DENSE method provides an optimal packing of strings into $2^m+1$ families. In principle, graph theory methods for the GC case could also find optimal or near-optimal solutions. As a practical matter, one should also consider the time/memory requirements of graph-based vs.\ our method.
Although an exhaustive study of graph-based GC algorithms is beyond the current scope,
for a comparison we tested the performance of the Largest First algorithm, as implemented in the {\tt rustworkx} package and provided through the Qiskit {\tt group\_commuting()} function. As a future study it would be interesting to compare with other heuristic methods in the literature. We found that for fully dense operators the DENSE method outperforms GC for $m \geq 2$, and that the ratio of families found by GC to that provided by DENSE increases with $m$. This is shown on the right side of Fig.~\ref{fig:nfam}. One may also consider the performance when the operator is not ``fully dense'', but instead contains some fixed percentage of the $4^m$ total strings.

The DENSE algorithm will produce a minimal number of cliques whenever $N_{\text{Pauli}} > 4^m-2^m$. When $N_{\text{Pauli}} \leq 4^m-2^m$, the solution is no longer guaranteed optimal, but may still be very good for sufficiently ``dense'' Hamiltonians. As $N_{\text{Pauli}}$ decreases, one would expect the relative GC performance to improve, since the dense method always finds $2^m+1$ cliques (except for the special case where a clique happens to be empty). We test this expectation in Fig.~\ref{fig:nfam}. We generate random Hamiltonians, and apply a cut on the Pauli string coefficients to reduce $N_{\text{Pauli}}$ prior to grouping the operators.
Note that for GC the result will depend on the detailed operator structure (i.e.\ the random operators tested here may not be representative of GC performance on other relatively dense operators of interest.) For example, on our tests of the Femtouniverse Hamiltonian~\cite{Butt:2022xyn}, which has a population of $\sim 50 \%$, the GC heuristic always finds $2^m$ families, outperforming DENSE. The results using random Hamiltonians are shown in Fig.~\ref{fig:nfam}. We find that for this class of operators, DENSE outperform GC until some (approximate) percentage threshold is reached, and this percentage decreases as $m$ increases. For $m=5$, DENSE outperforms until $N_{\text{Pauli}} \approx 0.2 \times 4^m$.

\begin{figure}
    \centering
    \includegraphics[width=0.49\textwidth]{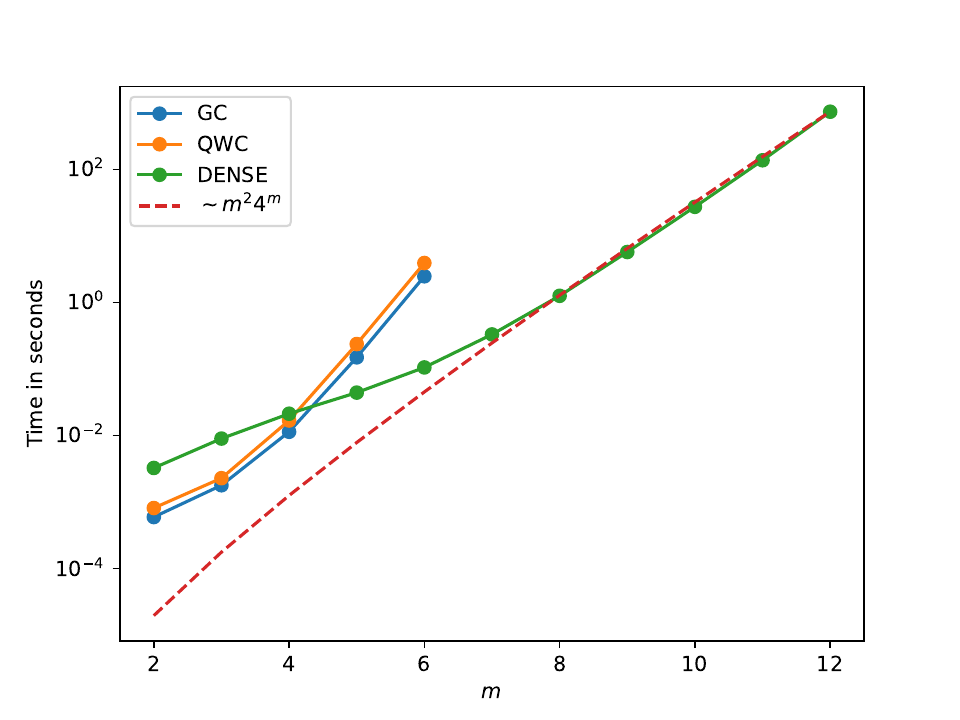}
    \includegraphics[width=0.49\textwidth]{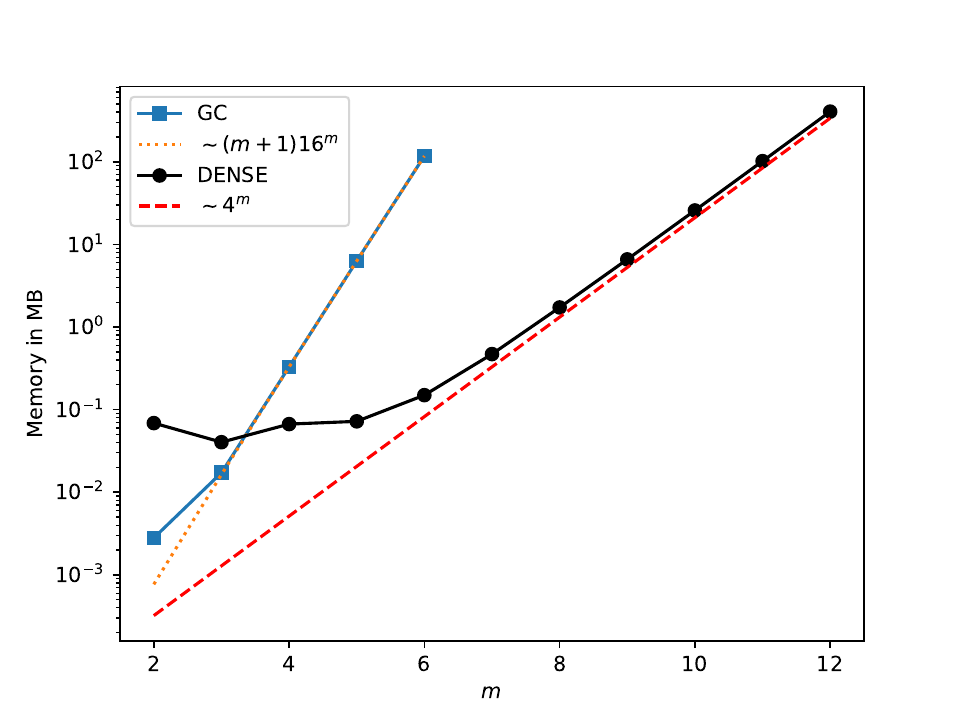}
    \caption{Walltime comparison to group random dense operators based on general (GC) and qubit-wise commuting (QWC) graph theoretic algorithms, and using methods based on finite fields (DENSE).}
    \label{fig:timing}
\end{figure}

We also investigated resource scaling of memory and walltime for the QWC, GC, and DENSE routines.
Note that these are classical resources used to generate solutions on a classical processor.
For graph-theoretic methods, an adjacency matrix is constructed which encodes the graph connectivity/commutation structure. For dense operators, this is a $4^m \times 4^m$ object,
and so the memory required will scale at least as $16^m$. We tested this expectation empirically again using the native {\tt Qiskit} implementation as a benchmark, shown in Fig.~\ref{fig:timing}.
Here we measured only the memory used in building the graph adjacency matrix (in the {\tt \_noncommutation\_graph()} subroutine); the peak memory usage was significantly larger as a vestige of casting numpy objects to python lists.
The rapid increase in memory usage meant that practically going beyond $m=6$ was not feasible on our laptops. In contrast, the DENSE routine generates families ``on the fly'' using powers of an $m \times m$ $A$ matrix, and the memory required to write down a solution will increase as $4^m$. 

We also compared walltimes for generating solutions.
For this comparison we simply timed calls to {\tt group\_commuting()} and {\tt PauliOrganizer()}.
For DENSE, solutions are generated either by 1) Computing the orbit of generators produced by matrix powers of $A$ or 2) enumerating the strings in the Pauli decomposition of an operator and using a lookup (detailed in~\cite{Reggio:2023fue}) to assign to a family. In either case the scaling will go like the number of strings $N_{\text{Pauli}}$ with sub-exponential corrections. Our measured walltimes are presented in Fig.~\ref{fig:timing}. For small $m$ the times to solution are comparable but as $m$ increases the favorable scaling of DENSE is evident. Note that the DENSE algorithm can also parallelize in a straightforward manner.

\section{Discussion}
We have presented a public code for grouping commuting Pauli strings, following a constructive algorithm which is optimal for observables that are dense in the space of Pauli strings. The algorithm is fast, and in terms of memory use and walltime outperforms public GC heuristics on random Hamiltonians and gauge-invariant matrix quantum mechanics models~\cite{Butt:2022xyn}. It would be interesting to develop applications to lattice gauge theory simulations, where using a local gauge invariant basis of states generally leads to dense subspaces with size that depends on the number of qubits dedicated to local regions of the lattice. It would also be interesting to explore hybrid dense/graph theoretic methods to make a  universally optimal solution to the grouping problem for arbitrary densities.
\section{Acknowledgements}

This work was supported in part by the U.S. Depart-
ment of Energy, Office of Science, Office of High Energy
Physics under award number DE-SC0015655 and by its
QuantISED program under an award for the Fermilab
Theory Consortium “Intersections of QIS and Theoretical Particle Physics.” We acknowledge the use of IBM
Quantum~\cite{ibmq} services for this work. The views expressed
are those of the authors, and do not reflect the official
policy or position of IBM or the IBM Quantum team


\end{document}